\documentclass[]{raa}            
\usepackage{graphicx,times}             

\usepackage{graphicx}
\usepackage{lscape}
\usepackage{longtable}

\begin{document}

\title{The study of Seyfert 2 galaxies  with and without infrared broad lines }

 \volnopage{ {\bf  } Vol.\ {\bf } No. {\bf XX}, 000--000}
   \setcounter{page}{1}

\author{Hong-bing Cai
      \inst{1,2}
   \and Xin-Wen Shu
      \inst{1}
   \and Zhen-Ya Zheng
      \inst{1}
   \and Jun-xian Wang
      \inst{1}
   }

   \institute{CAS Key Laboratory for Research in Galaxies and Cosmology, Department of Astronomy, University of Science and Technology of China, Hefei, Anhui 230026, P. R. China\\
     \and
              National Time Service Center, CAS, Xi'an, Shanxi,
             710600, China; {\it hbcai@ntsc.ac.cn}\\
\vs \no
   {\small Received [2009] [December] [17]; accepted [2010] [February] [23] }
}

\abstract{From the literature, we construct from literature a sample
of 25 Seyfert 2 galaxies (S2s) with a broad line region detected in
near infrared spectroscopy and 29 with NIR BLR which was detected.
We find no significant difference between the nuclei luminosity
(extinction-corrected [OIII]~5007) and infrared color
$\rm{f_{60}/f_{25}}$ between the two populations, suggesting that
the non-detections of NIR BLR could not be due to low AGN luminosity
or contamination from the host galaxy. As expected, we find
significantly lower X-ray obscurations in Seyfert 2s with NIR BLR
detection, supporting the unification scheme. However, such a scheme
was challenged by the detection of NIR BLR in heavily X-ray obscured
sources, especially in six of them with Compton-thick X-ray
obscuration. The discrepancy could be solved by the clumpy torus
model and we propose a toy model demonstrating that IR-thin
X-ray-thick S2s could be viewed at intermediate inclinations, and
compared with those IR-thick X-ray-thick S2s. We note that two of
the IR-thin X-ray-thick S2s (NGC 1386 and NGC 7674) experienced
X-ray transitions, i.e. from Compton-thin to Compton-thick
appearance or vice versa based on previous X-ray observations,
suggesting that X-ray transitions could be common in this special
class of objects.
 \keywords{galaxies: active -- galaxies: Seyfert -- infrared:
galaxies  } }

   \authorrunning{Cai et al. }            
   \titlerunning{Seyfert~2s with and without infrared broad lines}  
   \maketitle

\section{Introduction}

Spectropolarimetric observations have detected hidden broad emission
line regions in many Seyfert 2 galaxies (i.e., Antonucci \& Miller
\cite{A85}). This kind of S2s can be unified with Seyfert~1 galaxies
(S1s) under the unification scheme (Antonucci \cite{A93}), in which
the key component is  a dusty torus surrounding the nucleus, and
different lines of sight (obscured by the torus or not) induce
different apparent properties between S1 and S2. Strong evidence
supporting the unified model also comes from X-ray observations,
which show that most S2s are generally heavily obscured (N$\rm{_H}>
10^{22}$~cm$^{-2}$) (Risaliti et al. \cite{R99}; Bassani et al.
\cite{B99}). However, only $\sim$ 50\% of S2s with
spectropolarimetric observations show polarized broad emission lines
(PBLs) in polarized spectra (e.g. Tran 2001). The visibility of PBLs
was found to be dependent on many factors, including intrinsic AGN
luminosity, infrared colors, nuclei obscuration, and likely more,
and provides valuable information on the physical properties and
geometry of the obscuration, scattering region, and the broad
emission line region (e.g., Shu et al. 2007 and references therein).

Another efficient way to detect the hidden broad line regions in S2s
is near-infrared (NIR) spectroscopy, due to the relatively smaller
dust opacity in NIR than in the optical (e.g. see Gordon et al. 2003
for a Galactic dust extinction curve). For instance, the Galactic
dust extinction at 4.05$\mu$m is $\sim$ 20 times weaker than in the
V band (A$_{4.05\mu{m}}$/A$_V$ = 0.051, Lutz 1999). NIR observations
have succeeded in detecting NIR broad emission lines in some (but
not all) S2s (e.g., Goodrich et al. \cite{G94}; Veilleux et al.
\cite{V97a}; Lutz et al. \cite{L02}), providing further support to
the AGN unification model.
By studying a sample of 12 type 2 AGNs, Lutz et al. (2002) found all
three objects with detected broad Br$\alpha$ exhibit relatively low
X-ray obscuring columns (N$\rm{_H}$ $<$ 10$^{23}$ cm$^{-2}$), while
those without are more heavily obscured in X-ray (with  N$\rm{_H}$
$>$ 10$^{23}$ cm$^{-2}$). More interestingly, the NIR opacities they
found were consistent with a Galactic ratio of 4 $\mu$m obscuration
of the BLR and X-ray column.

However, careful examinations of other factors which could affect
the visibility of NIR BLR have not been performed, and better
understanding of the relation between NIR and X-ray obscuration in
Seyfert 2 galaxies demands larger samples. In this paper, we
construct a large sample of S2s from literature with detected NIR
broad emission lines (IRBL S2 hereafter) and those which were not
detected (non-IRBL S2), and perform the first systematic comparison.
Throughout this paper, we use the cosmological parameters H$_0
=$70~km~s$^{-1}$~Mpc$^{-1}$, $\Omega\rm{_m} =$0.27, and $\Omega
\rm_{\lambda} =$0.73.


\section[]{Sample and Statistical
Properties}

By searching the literature, we collected 25 IRBL S2s. We exclude
intermediate Seyferts (Seyfert 1.8, Seyfert 1.9) which will
otherwise bias our comparison of obscuration (see \S 3) since they
were defined to be less obscured in the optical, and thus most of
them are observable in the near infrared and X-ray.
Most of these infrared broad lines are hydrogen recombination lines,
i.e., Bracket and Paschen lines, except for one He~I~10830 in
NGC~1275. The sample and the detected NIR broad emission lines and
widths are listed in Table 1. We also list the following in the
table for each source: in Cols.~(5-6) whether or not polarized broad
emission lines were detected and their corresponding references; in
Col.~(7) IR flux density ratio  $f\rm_{60\mu m}$/$f\rm_{20\mu m}$,
where $f\rm_ {60}$ and $f\rm_{20}$ are directly from NED
except for 
Mrk~477 and 3C~223 whose values are
interpolatively or extrapolatively obtained according to their
infrared spectrum; column (8-9) X-ray absorption column density and references;
column (10-12) reddening-corrected [O III] line flux, luminosity and references.
For comparison, we list the sample of 29 non-IRBL S2s in Table~2.
LINERs are excluded from the sample.

When calculating the luminosity of \rm{[O~III]} in Table~1 and
Table~2, we have applied  $\rm{F^{cor}_{[O~III]}}= \rm{F^{obs}_
{[O~III]}[(H\alpha/H\beta)_{obs}/(H\alpha/H\beta)_0]}^{2.94}$
(Bassani et al. \cite{B99}) to correct the dust extinction to
\rm{[O~III]}, where $\rm{F^{cor}_{[O~III]}}$ is the
extinction-corrected flux of [O~III]~5007,  and $\rm{F^{obs}_
{[O~III]}}$ the observed flux of \rm{[O~III]}~5007 with intrinsic
Balmer decrement of $\rm{(H\alpha/H\beta)_0}=$3.0.

\begin{center}
\setlength{\tabcolsep}{0.8pt} \footnotesize
\begin{longtable}{cccccccccccc}

\caption{Sample of Seyfert~2s with near-infrared broad lines.}\\
\hline

Source   & z & Width& Ref. &  PBL & Ref. &  $f_{60}/f_{25}$ & $\log
\rm{N_H}$
&Ref.
  & F$_{\rm{[O~III]}}$&logL$_{\rm{[O~III]}}$  & ref. \\
(1) & (2) & (3) & (4) & (5) & (6) & (7) & (8) & (9) & (10) & (11)&
(12)\\

\hline
\endfirsthead
\hline

Source   & z & Width& Ref. &  PBL & Ref. &  $f {60}/f {25}$ & $\log
\rm{N H}$ & Ref.
  & F$ {\rm{[O~III]}}$&logL$ {\rm{[O~III]}}$ & ref. \\
(1) & (2) & (3) & (4) & (5) & (6) & (7) & (8) & (9) & (10) & (11)&
(12)\\
\hline
\endhead
\hline

3C~223 & 0.14&Pa$\alpha$4100&H96&?&\ldots &0.26  & 22.88 &C04&3.88&42.28&P07\\
F13451+1232 & 0.12&Pa$\alpha$2588&V97b&? &\ldots& 2.88 & 22.65&I04&2.79&42.02&A00\\
F23498+2423  & 0.21&Pa$\alpha$3027&V97b&?&\ldots & 8.51 & \ldots&\ldots &12.11&43.2& V99\\
F13305-1739  & 0.148&Pa$\alpha$2896&V99&? &\ldots& 2.95 &\ldots&\ldots&361.96&44.33&V99\\
IRAS~08311-2459 & 0.10&Pa$\alpha$779&M00&?&\ldots & 4.53 &\ldots &\ldots&\ldots&\ldots &\ldots\\
IRAS~15462-0450 & 0.10&Pa$\alpha$&M99&?&\ldots & 6.48 &\ldots&\ldots&6.15&42.19&V99\\
NGC~6240 &0.024&Br$\alpha$1800&S08& ? &\ldots& 6.48 &  24.13&P03&135&42.27&B99\\
 Arp~220  & 0.018&Br$\alpha$3300&D87&?&\ldots & 13.01 & 22.48 &C02&3.69&40.44&V99\\
 IRAS~14348-1447 & 0.083&Pa$\alpha >$2500&N91&?&\ldots & 11.48 & 21.48&F03&4.53&41.89&V99\\
 NGC~1386     & 0.0029&Br$\gamma$&R02&n&M00 & 3.74 &  $>$24&L06&1020&41.27& S89 \\
 NGC~7582 & 0.0052&Br$\gamma$3000&R03&n &H97& 6.62 & 23.89 &P07&369&41.36&S95\\
 Mrk~463E  & 0.05&Pa$\beta$1794,Br$\gamma$1070&V97a&y&Y96 & 1.35 &  23.51&I04&125&42.89&D88\\
 Mrk~477  & 0.038&Br$\gamma$1630&V97a&y&T92  &  2.71 &  $>$24&B99&1240&43.62&D88\\
 NGC~2110  & 0.0078&Pa$\beta$1204,Br$\gamma$3127&V97a&y&M07 & 4.92  &  22.48&E07&321&41.62&B99\\
 NGC~262, Mrk~348   & 0.015&Pa$\beta$1850,Br$\gamma$2500&V97a&y&D88 & 1.54  &  23.20&A00&177&41.96&D88 \\
 NGC~7674  & 0.029&Pa$\alpha$3000,Br$\gamma$3000&R06&y&Y96 &  2.79 &  $>$24&B05a&193&42.57&D88 \\
 F20460+1925  & 0.18&Pa$\alpha$2857&V97b&y&Y96 & 1.55 & 22.40&B99&11.2&43.02&B99\\
 F23060+0505  & 0.17&Pa$\alpha$1954&V97b&y&D04 & 2.69 & 22.92&B07&99.45&43.92&B99 \\
 F05189-2524  & 0.042&Pa$\alpha$2619&V99&y&Y96 & 3.98 &  22.83&T09&130&42.74&L01\\
 IRAS 11058-1131  & 0.055&Pa$\alpha$7179&V99&y&Y96 & 2.4  & $>$24&U00&39.4&42.45&Y96  \\
 Mrk~3        & 0.014&\ldots&H99&y&M90 & 1.34 & 24.13&B05b&4610&43.27&M94 \\
 Mrk~176 & 0.027&Pa$\beta$1329&R94&y&M83 & 2.96 &  23.88&G05&105&42.19&K78\\
 PKS~1549-79 & 0.15&Pa$\alpha$1745&B03&y&S95 & 2.15 & \ldots&\ldots&130&41.15&H06\\
 3C~234 &  0.184&Pa$\alpha$4100&H96&y&Y98  & 0.89 &  23.54&Pi08&20.9&43.3&P08 \\
NGC~1275 & 0.018&He~I 1.08$\mu$m4700&R06&y&G01 & 2.11 & 21.08 &C03&311&42.34&B99\\

\hline

\end{longtable}

\tablecomments{1.0\textwidth}{ Col. (1): the source name; col. (2):
the redshift; col. (3): the width of NIR broad lines. The letters
and number denote the type of lines and their velocities (in unit of
km~s$^{-1}$), respectively; col. (4): the references for the
observations of near-infrared broad lines; col. (5): whether have
been observed optically polarized broad lines; col. (6): the
references for the observations of polarized broad lines; col. (7):
the ratio of the flux at 60$\mu m$ to the flux at 25$\mu m$; col.
(8), (9): the hydrogen density by X-ray observations and their
references, respectively; col. (10), (11), (12): the redden
corrected \rm{[O~III]} flux density (in units of
10$^{-14}$~erg~cm$^{-2}$~s$^{-1}$), \rm{[O~III]} luminosity (in
units of erg~s$^{-1}$), and their references, respectively.\\
\vspace{-1mm}\\ Ref. A00: Axon 2000; A06: Awaki et al. 2006; B99:
Bassani et al. 1999; B03 Bellamy et al. 2003; B07: Bian et al. 2007;
B05a: Bianchi et al. 2005a; B05b: Bianchi et al. 2005b; C02:
Clements et al. 2002; C03: Churazov et al. 2003; C04: Croston et al.
2004; D88: Dahari \& De Robertis 1988;
D04: Deluit 2004; D87: DePoy et al. 1987; 
E07: Evans et al. 2007; F03: Franceschini et al. 2003; G01: Gu et
al. 2001; G05: Guainazzi et al. 2005; H97: Heisler et al. 1997; H99:
Heisler \& Robertis, 1999; H96: Hill, et al. 1996;H06: Holt et al.
2006; I04: Imanishi \& Terashima et al. 2004; K78: Koski et al.
1978; L06: Levenson et al. 2006; L01: Lumsden et al. 2001; L04:
Lumsden et al. 2004; L02: Lutz et al. 2002; M83: Martin et al. 1983;
M90: Miller \& Goodrich 1990; M00: Moran et al. 2000; M94: Mulchaey
et al. 1994; M99: Murphy et al. 1999;
M00: Murphy et al. 2000; N91:
Nakajima et al. 1991; N05:
Netzer 2005; O00: 
P03: Ptak et al. 2003; P07: Piconcelli et al. 2007; Pi08: Piconcelli
et al. 2008; P08: Privon et al. 2008; Reunanen et al. 2002; R03:
Reunanen et al. 2003; R06: Riffel et al. 2006; R94: Ruiz et al.
1994; S08: Sani et al. 2008; S95: Shaw et al. 1995; S89:
Storchi-Bergmann \& Pastoriza 1989; T09:Teng et al. 2009; T01: Tran
2001; 
U00: Ueno et al. 2000; V97a: Velilleux et al. 1997a; V97b: Velilleux
et al. 1997b; V99: Veilleux et al. 1999; W93: Warwick et al.
1993,MNRAS; Y96: Young et al. 1996; Y98: Young et al. 1998.
 }
\end{center}

\begin{center}
\setlength{\tabcolsep}{1.8pt} \footnotesize
\begin{longtable}{cccccccccccc}
\caption{Sample of Seyfert~2s without the detection of near-infrared broad lines.}\\
\hline

Source & z & Ref. & PBL & Ref. &  $f_{60}/f_{25}$ & $\log \rm{N_H}$
& Ref. & F$_{\rm{[O~III]}}$ &logL$_{\rm{[O~III]}}$ & Ref.\\ 
(1) & (2)  & (3) & (4) & (5) & (6)  & (7) & (8) & (9) & (10) & (11)\\

\hline
\endfirsthead
\hline

Source & z & Ref. & PBL & Ref. &  $f_{60}/f_{25}$ & $\log \rm{N_H}$

& Ref.
 & F$_{\rm{[O~III]}}$ &logL$_{\rm{[O~III]}}$ &Ref. \\
(1) & (2)  & (3) & (4) & (5) & (6)  & (7) & (8) & (9) & (10) &
(11)\\

\hline
\endhead
\hline

 Mrk~273 & 0.038&V99&?&\ldots & 9.55 & 23.61&X02 &84& 42.44&B99\\
 Mrk~1073    & 0.023&V97a&?&\ldots & 5.71 & $>$24&G05 &21.95&41.42&B97\\
 F12072-0444 & 0.13&V97b&?&\ldots & 4.57 & $>$24 &T05&118.35&43.71&V99 \\
 Mrk~622     & 0.023&V97a&?&\ldots & 1.28 &  $>$24&G05&3.92&40.68&B97    \\
 NGC~4968 & 0.0099&L02&?&\ldots & 2.27 & $>$24&L02&1116&42.38& B99\\
 NGC~7172 & 0.0087&V97a&n &H97& 6.64 & 22.95&S07 &4.04&39.83&Va97\\
 NGC~5643 & 0.0040&L02&n&M00 & 5.16 &  $>$25&L02 &661.76&41.37&M94\\
 NGC~1142 & 0.029&R06&n&T01 & 8.37 & 23.65&Sa07 &38.82&41.87&V95\\
 NGC~7130 & 0.016&V97a&n&H97 & 7.76 &   $>$24&L05 &600&42.55&S95\\
 NGC~34   & 0.020&R06&n&H97 & 6.96 &    $>$24&S07&768&42.83&V95 \\
 Mrk~266  & 0.028&V97a&n&T01 & 4.6  &  $>$25&R00  &44&41.89&D88\\

 NGC~4941 &0.0037&L02& n&M00 & 2.61 & 23.65&L02&457&41.14&S89\\
 ESO~428-G14 & 0.0057&V97a&n&M00 & 2.49 & $>$24&L06 &2010&42.15&A91\\
 Mrk~78 & 0.037&V97a&n&M90 & 2 & ... &\ldots&105.7&42.53&K78 \\
 Mrk~1   & 0.016&V97a&n&K78 & 2.93 &   $>$24&G05 &55.04&41.5&N00\\
 NGC~4945 &0.0019&R02& n&M06 & 14.17 &  24.70&I08&$>$40& $>$39.48&R99\\
 NGC~5728 & 0.0094&V97a&n&M00 & 9.24 & 23.91&Z06&761&42.18&S95\\
 Mrk~1066 & 0.012&V97a&n&M00 & 4.77 & 23.95&G05&514&42&M94\\
 F04103-2838 & 0.12&V99&n&Y96 & 3.39 & $>$24&T08&21.34&42.88& V99\\
 NGC~5128 &0.0018&R02& n&A99  & 8.47 & 23.00&E04 &$>$7&$>$38.73&B99\\
 NGC~3081 &0.007976&V97a& y&M00 &\ldots  & 23.82&B99&195&41.43& S95\\
 NGC~5929 &0.0083&R06& y&M01 & 5.64 & ... &S07&153&41.4& L01\\
 NGC~4388 & 0.0084&V97a&y&Y96 & 2.7 & 23.32&C06&451&41.85&D88\\
 NGC~1068 & 0.0038&V97a&y&A85 & 2.12 & $>$24&P06 &6780&42.33&D88\\
 Mrk~573 &0.017&V97a& y&N04 & 1.34 & $>$24&G05 &177&42.08&D88\\
 Circinus & 0.0014&L02&y&O98 & 3.63 & 24.60&L02&1848.6&40.92&L05\\
 NGC~7212 & 0.027&V97a&y&T92 & 3.75 & $>$24&G05 &320&42.73&M94 \\
 Mrk~1157 & 0.015&V97a&y&M00 & 4.44 &  $>$24&G05 &178&41.97& M94 \\
 Mrk~1210  & 0.014&L02&y&T92  &  0.91 & 23.26&O04&580&42.37& T91\\

\hline
\end{longtable}
\tablecomments{0.9\textwidth}{ Col. (1): the source name; col. (2):
the redshift; col. (3): the references for the observations of
near-infrared broad lines; col. (4): whether have been observed
optically polarized broad lines; col. (5): the references for the
observations of polarized broad lines; col. (6): the ratio of the
flux at 60$\mu m$ to the flux at 25$\mu m$; col. (7), (8): the
hydrogen density by X-ray observations, and their references,
respectively; col. (9), (10), (11): the redden corrected
\rm{[O~III]} flux density (in units of
10$^{-14}$~erg~cm$^{-2}$~s$^{-1}$), \rm{[O~III]} luminosity (in
units of erg~s$^{-1}$), and their references, respectively.\\
\vspace{-0.5mm}\\Ref. A91: Acker et al. 1991; A99: Alexander et al.
1999; A85: Antonucci\& Miller 1985; B99: Bassani et al. 1999; B97:
Bonatto et al. 1997; C06: Cappi et al. 2006; D88: Dahari \& De
Robertis 1988; D04: Deluit 2004; D03: Dennefeld et al. 2003;
E04: Evans et al. 2004; G05: Guainazzi et al. 2005; H97: Heisler et
al. 1997; I08: Itoh et al. 2008, PASJ; K78: Koski 1978; L05:
Levenson et al. 2005; L06: Levenson et al. 2006; L01: Lumsden et al.
2001; L02: Lutz 2002; M06: Madejskiet al. 2006; M00: Moran etal.
2000; M01: Moran 2001; M94: Mulchaey et al. 1994; N00: Nagao et al.
2000; N04: Nagao et al. 2004; O04: Ohno et al. 2004; O98: Oliva et
al. 1998; 
P06: Pounds 2006; R06: Riffel et al. 2006; R99: Risaliti et al.
1999; R00: Risaliti et al. 2000; Sa07: Sazonov et al. 2007; S07: Shu
et al. 2007; S89: Storchi-Bergmann \& Pastoriza 1989; S95:
Storchi-Bergmann et al. 1995; T05: Teng et al. 2005; T08: Teng et
al. 2008; T91: Terlevich et al. 1991; T92: Tran et al. 1992; T01:
Tran et al. 2001; Va97: Vaceli et al. 1997; V95: Veilleux et al.
1995; V97a: Veilleux et al.1997; V97b: Veilleux et al. 1997; V99:
Veilleux et al. 1999; X02: Xia et al. 2002; Y96: Young et al. 1996;
Z06: Zhang J.S 2006. }


\end{center}


In Figure 1 we plot the redshift distribution of both IRBL S2s and
non-IRBL S2s. Most of the sources are located at redshift $<$ 0.05, except for several IRBL S2s.
K-S test shows that the redshift distributions of IRBL S2s and
non-IRBL S2s are statistically different with a confidence level of
99.8\%. We attribute this difference to the inhomogeneous nature of
the samples (combined from literature).

\begin{figure}
\begin{center}
 \scalebox{0.6}{\includegraphics{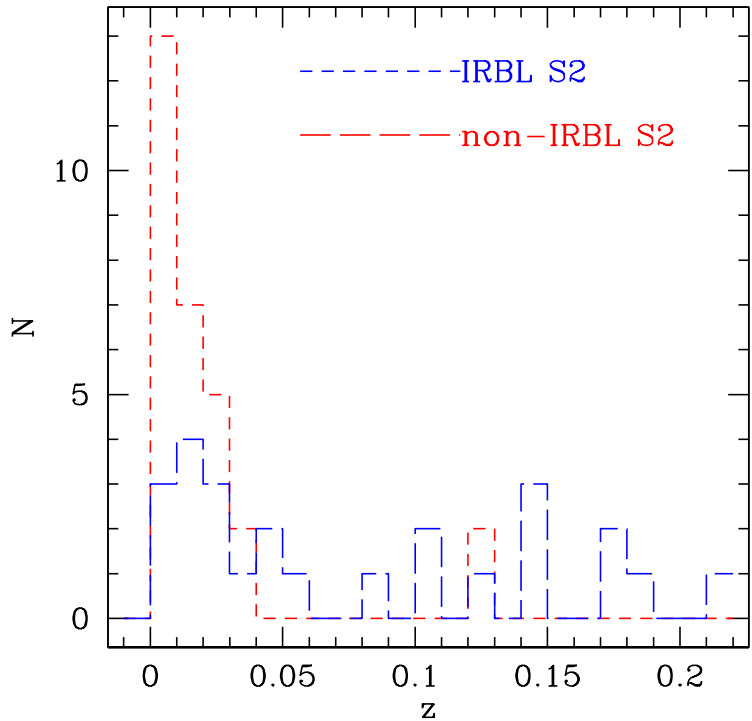}}
 \caption{The distribution of redshifts of
IRBL S2s and non-IRBL S2s.}
\end{center}
\end{figure}

Various studies had found higher detection rate of PBLs in S2s with
higher AGN luminosity (see Shu et al. 2007 and references therein).
To investigate whether the visibility of NIR broad emission lines in
S2s depends on AGN's intrinsic luminosity, we plot in Fig. 2 the
histogram distribution of reddening corrected L$_{\rm{[O~III]}}$ for
IRBL S2s and non-IRBL S2s. We find a weak trend that non-IRBL S2s
show lower L$_{\rm{[O~III]}}$, however the difference is
statistically insignificant (K-S test shows the difference has only
a confidence level of 90\%).

\begin{figure}
\begin{center}
\scalebox{0.6}{\includegraphics{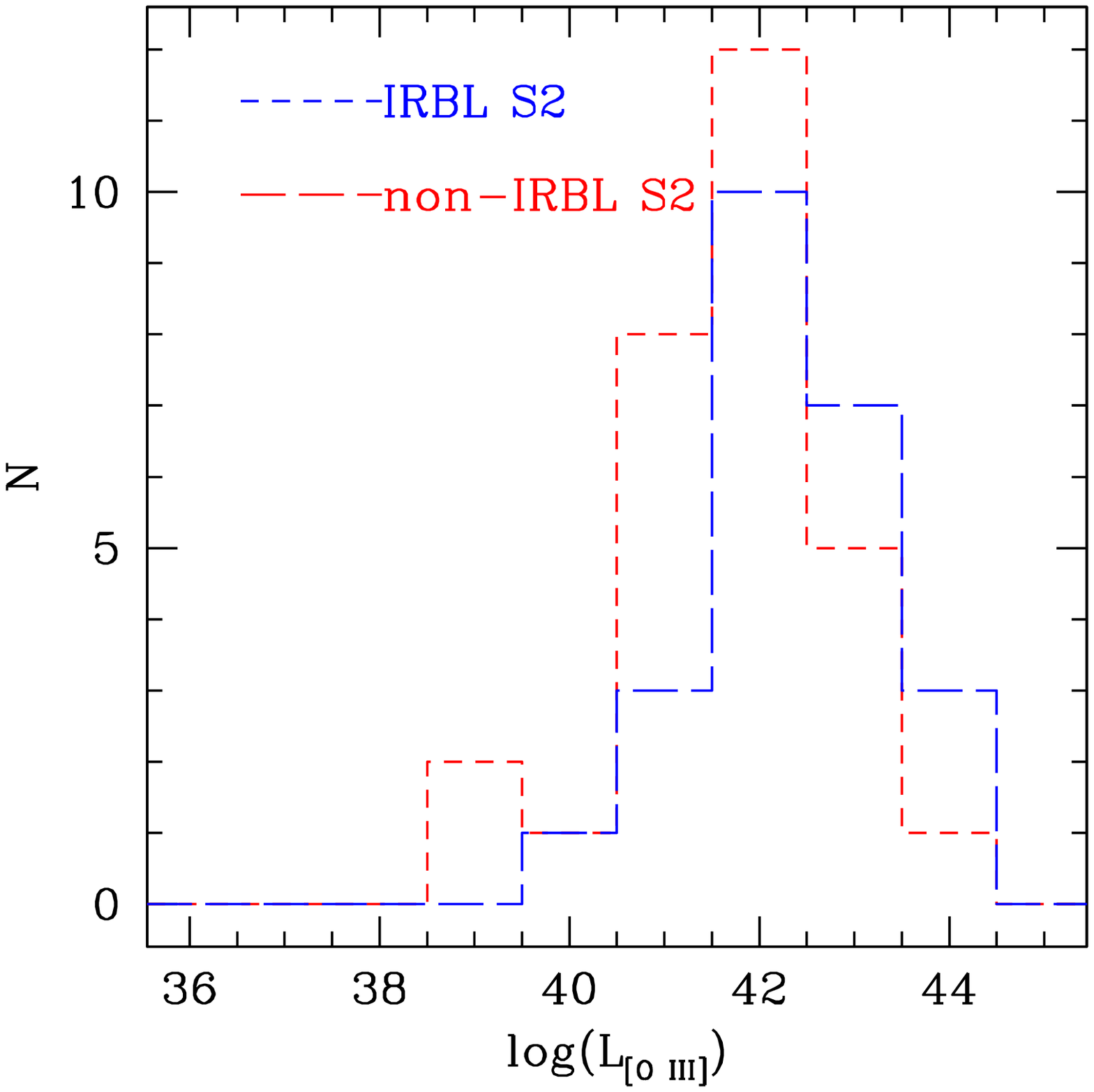}}
 \caption{The distribution of L$_{\rm{[O~III]}}$ of
IRBL S2s and non-IRBL S2s.}
\end{center}
\end{figure}

Heisler et al. (\cite{H97}) found that the visibility of PBL is
related to the far-infrared colors and Seyfert~2s with warmer FIR
colors ($f_{60}/f_{25} <4$) are more likely to have hidden broad
lines. Here we also investigate whether the visibility of NIR broad
lines in S2s depends on FIR colors. We plot the distributions of FIR
colors (f$_{60\mu m}$/f$_{25\mu m}$) of IRBL S2s and non-IRBL S2s in
Fig.~3,
and find no statistical difference between two populations.

\begin{figure}
\centering \scalebox{0.6}{\includegraphics{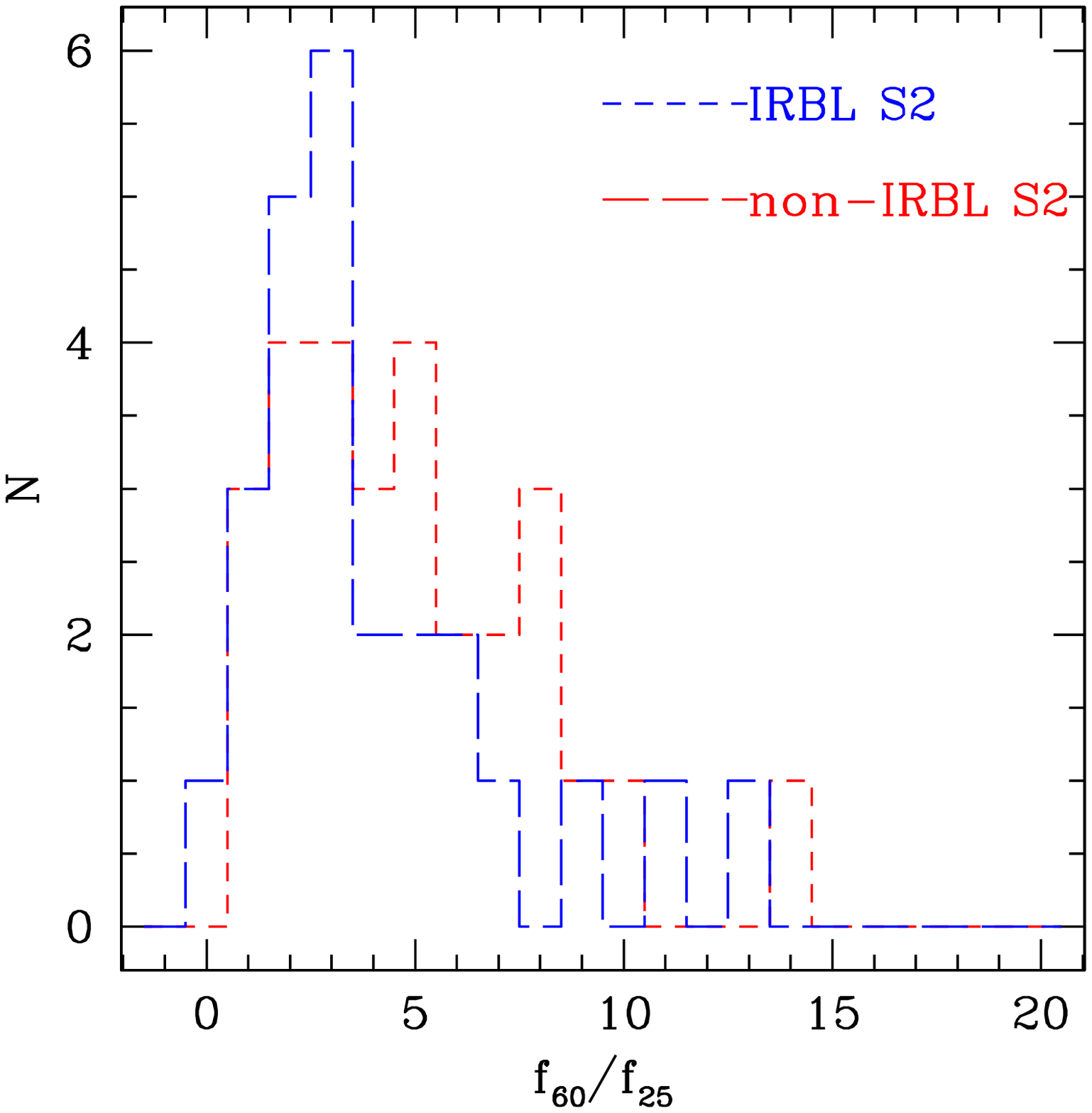}} \caption{The
distributions of $f_{60}/f_{25}$ of IRBL S2s and non-IRBL S2s.}
\end{figure}

The X-ray absorption column densities were collected from literature
for two samples. Whenever available, we cite measurements based on
Chandra/XMM data which have higher spatial resolution and higher
spectral quality comparing with previous data. In Fig.~4 we plot the
N$\rm{_H}$ distribution for IRBL S2 and non-IRBL S2, where we can
clearly see weaker X-ray obscuration in IRBL S2s. K-S test shows
that the difference in N$\rm{_H}$ between two populations is
significant at 99\% level.

\begin{figure}
\begin{center}
\scalebox{0.6}{\includegraphics{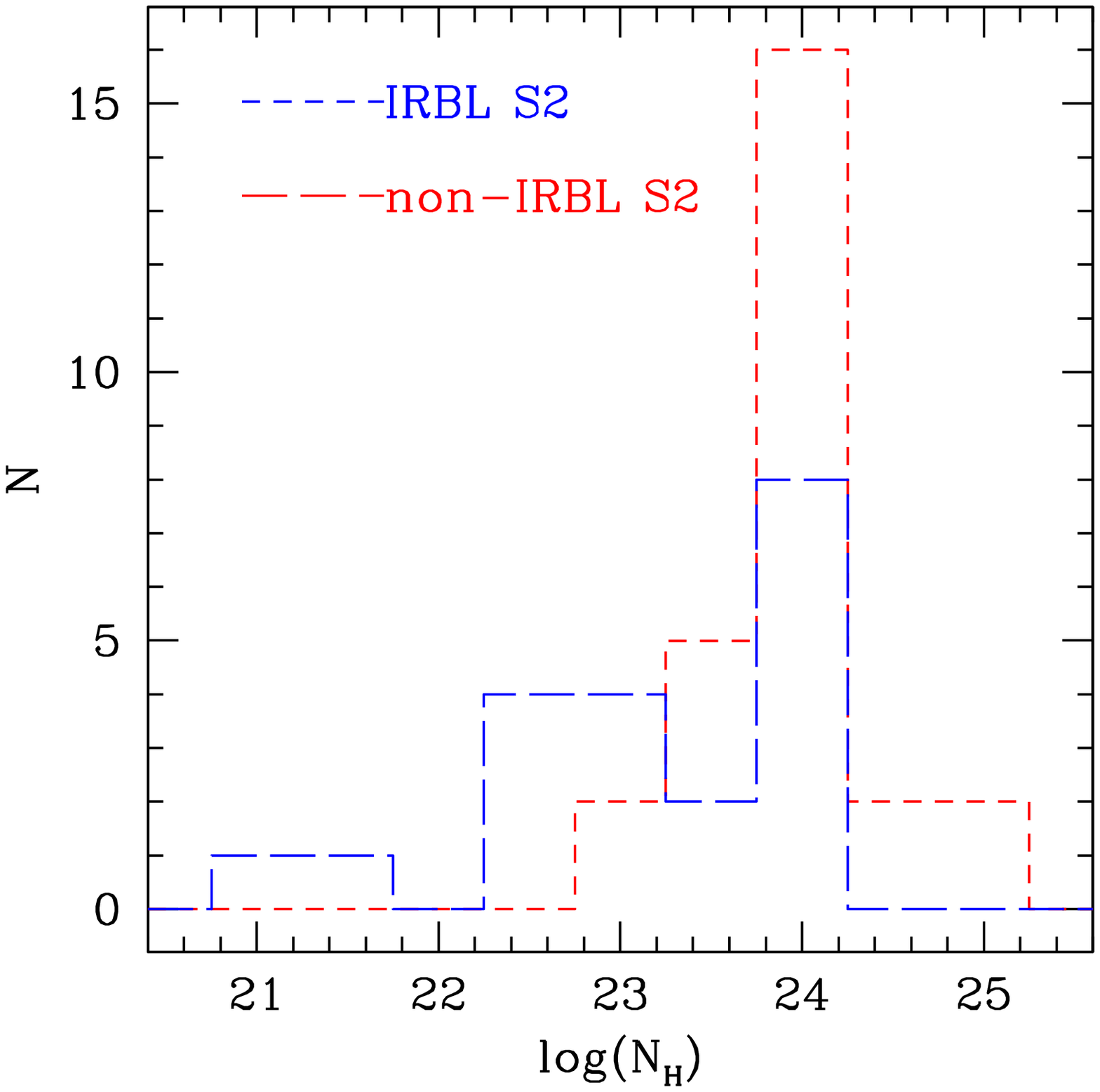}} \caption{The
distribution of the hydrogen column densities of IRBL S2s  and
non-IRBL S2s. }
\end{center}
\end{figure}

\section{Discussion}

\subsection{IRBL S2s versus non-IRBL S2s}


Comparing our samples of IRBL S2s with non-IRBL S2s, we find that while IRBL S2s have significantly higher redshift, their intrinsic AGN luminosity distribution is statistically consistent with non-IRBL S2. The consequence is that
the non-detection of infrared broad emission lines in non-IRBL S2s,
which are located at even lower redshifts,  is not likely due to the
detection limit of IR spectral observations. The similar
distribution of FIR colors ($f_{60}/f_{25}$) in two populations and
the smaller redshift in non-IRBL S2s (which suggests smaller
contribution from the host galaxy for fixed slit width) also suggest
that the non-detection of NIR broad emission lines could not be
attributed to stronger host galaxy contamination.

Recent studies have suggested that the BLR might be absent in low luminosity and/or low accretion rate AGNs (e.g., Elitzur \& Ho et al. 2009). However, the consistent intrinsic AGN luminosity in IRBL S2s and non-IRBL S2s suggest that the non-detection of NIR BLR in our sample was not due to intrinsic absence of BLR, but due to heavy obscuration.

Consistently, we find significantly higher X-ray obscuration in non-IRBL S2s, 66\% of which are obscured with column density $>$ 10$^{24}$ cm$^{-2}$, while the fraction for IRBL S2s is only 29\%. This result is consistent with Lutz et al. (2002), which detected low X-ray absorption  in IRBL S2s in a much smaller sample.


We also check the relation between the detection of NIR broad emission line and that of polarized broad emission line in Seyfert 2 galaxies. By matching our sample with spectropolarimetric observations in literature (see Table 1 and 2), we find that PBLs were detected in 14 out of 16 IRBL S2s with both NIR spectral and optical spectropolarimetric observations,  but in a much smaller fraction (9 out of 24) of non-IRBL S2. This suggests that the visibility of BLR in NIR spectra and in polarized spectra is connected. Shu et al. (2007) found S2s with PBL have smaller X-ray obscuration. This pattern is consistent with our finding that S2s with NIR broad emission lines have lower X-ray obscuration.

\subsection{Dust extinction versus X-ray obscuration}

The optical dust extinction in AGNs (E$\rm{_{B-V}}$  or A$\rm{_V}$)
has been compared with X-ray obscuring column densities by many
studies, which yield rather large diversity in the observed
dust-to-gas ratios.  Particularly, Maiolino et al. (2001a) reported
E$\rm{_{B-V}}$/N$\rm{_ H}$ appears $\sim$ 3 to ~$\sim$ 100 times
lower than Galactic in various class AGNs, most of which are Seyfert
1s, quasars and intermediate Seyferts. Maiolino et al. (2001b)
attributed the lower E$\rm{_{B-V}}$/N$\rm{_ H}$ to several possible
mechanisms, including dust distribution dominated by large grains
(their most favorable model), metallicity higher than solar which
would affect N$\rm{_ H}$ measurements through X-ray spectral
fitting, low dust-to-gas ratio, geometry effect, etc (also see Wang
\& Zhang 2007). On the other hand, dust-to-gas ratio consistent with
or even much higher than Galactic value were also reported. For
instance, Wang et al. (2009) reported a dust-to-gas ratio consistent
with Galactic value in the partially obscured Seyfert galaxy
Mrk~1393, while much higher dust-to-gas ratios were reported in some
AGNs which could be due to ionized absorber mixed with the dust (see
Komossa 1999 for a review).

Taking the large diversity in the ratio of observed dust reddening
to X-ray column density, we are not surprised to see significant
number of IRBL S2s with large X-ray obscuration (e.g. those with
N${\rm_H}$ $>$ 10$^{23}$ cm$^{-2}$ in Fig. 4), which were absent
from the small sample of Lutz et al. (2002). The possible mechanisms
proposed by Maiolino et al. (2001b) could also apply here. We note
that the samples we presented are not uniformly selected, since
various NIR observations were designed to search for various NIR
broad emission lines with different detection limit. A more strict
comparison between two populations requires direct measurements of
NIR extinction for each detected NIR broad emission line, and lower
limits to those non-detected, most of which were unavailable however
from literature.

However, the detection of IR broad emission lines in those X-ray
Compton-thick S2s (with N$\rm_H$ $>$ 10$^{24}$ cm$^{-2}$) is still a
puzzle, since the thick X-ray absorption is too large for the
detection of NIR BLR (e.g. the detection of Br$\alpha$ implying
N${\rm_H}$ $<$ 10$^{23}$ cm$^{-2}$ assuming a Galactic extinction
curve, see Lutz et al. 2002). The discrepancy can not be solved even
assuming a dust-to-gas ratio 10 times lower than Galactic. Dust
dominated by large grains could not help either since dust with
large grains has a much flatter extinction curve, and  does not
significantly alter the opacity in NIR (see Fig. 5 of Lutz et al.
2002).


\subsection{IR-thin X-ray-thick S2s and their Spectral transitions}

For six of IRBL S2 identified as Compton-thick in X-ray (NGC 1386, NGC 7674, Mrk~3, NGC~6240, Mrk~477, and IRAS 11058-1131), we refer them as IR-thin X-ray-thick Seyfert 2 galaxies hereafter.
Below we discuss several  possible explanations to the discrepancy between the detection of NIR broad lines and Compton-thick X-ray obscuration.

In the view of traditional unification model, this discrepancy can be ascribed to the different emission region of broad emission line and X-ray continuum. These sources could be viewed along the edge of the torus where the absorption along the line of sight  to the central engine is still X-ray thick, while that to the outer part of BLR at larger distances has much lower column density. In this scheme, the column density of the torus smoothly decreases from edge-on to face-on, and these sources (infrared-thin X-ray thick) can only be detected at intermediate inclination angle.

However, the geometry of the torus in AGN is still unclear.  The recent view of the
unification model has suggested that the obscuring torus is most likely clumpy,
instead of smoothly distributed. The evidences come for instance from the IR SED fitting  (e.g. Ramos Almeida et al. 2009 ), and the detection of rapid
variation of X-ray absorption in Seyfert 2 galaxies (e.g. Risaliti et al. 2007).  In this model, it's much easier
to understand the observed different opacities detected in IR and X-ray band: we could
 expect that BLR is mainly obscured by Compton-thin medium, with clumpy Compton-thick
clouds mixing within it. The typical size of a single Compton-thick
cloud must be much smaller than BLR, otherwise one would see
consistent N$\rm{_H}$ inferred from optical/NIR and  from X-ray.
Nevertheless, considering that NIR BLR have been detected only in a
small fraction of Compton-thick S2s, IR-thin/X-ray-thick S2s must be
viewed differently from those IR-thick/X-ray-thick ones. Comparing
with IR-thick X-ray-thick sources, they could be at certain evolving
stage or at intermediate inclination angle  when/where the clumpy
Compton-thick clouds do exist to block the central X-ray continuum
with certain probability, but with too small filling factors to
block the whole BLR. In Fig. 5 we present a cartoon diagram to
demonstrate the toy model we proposed. If the model is right, we
would expect X-ray absorption variations in these IR-thin X-ray
thick Seyfert 2 galaxies, due to the movement of the compact
Compton-thick clouds.

\begin{figure}
\begin{center}
\scalebox{0.8}{\includegraphics{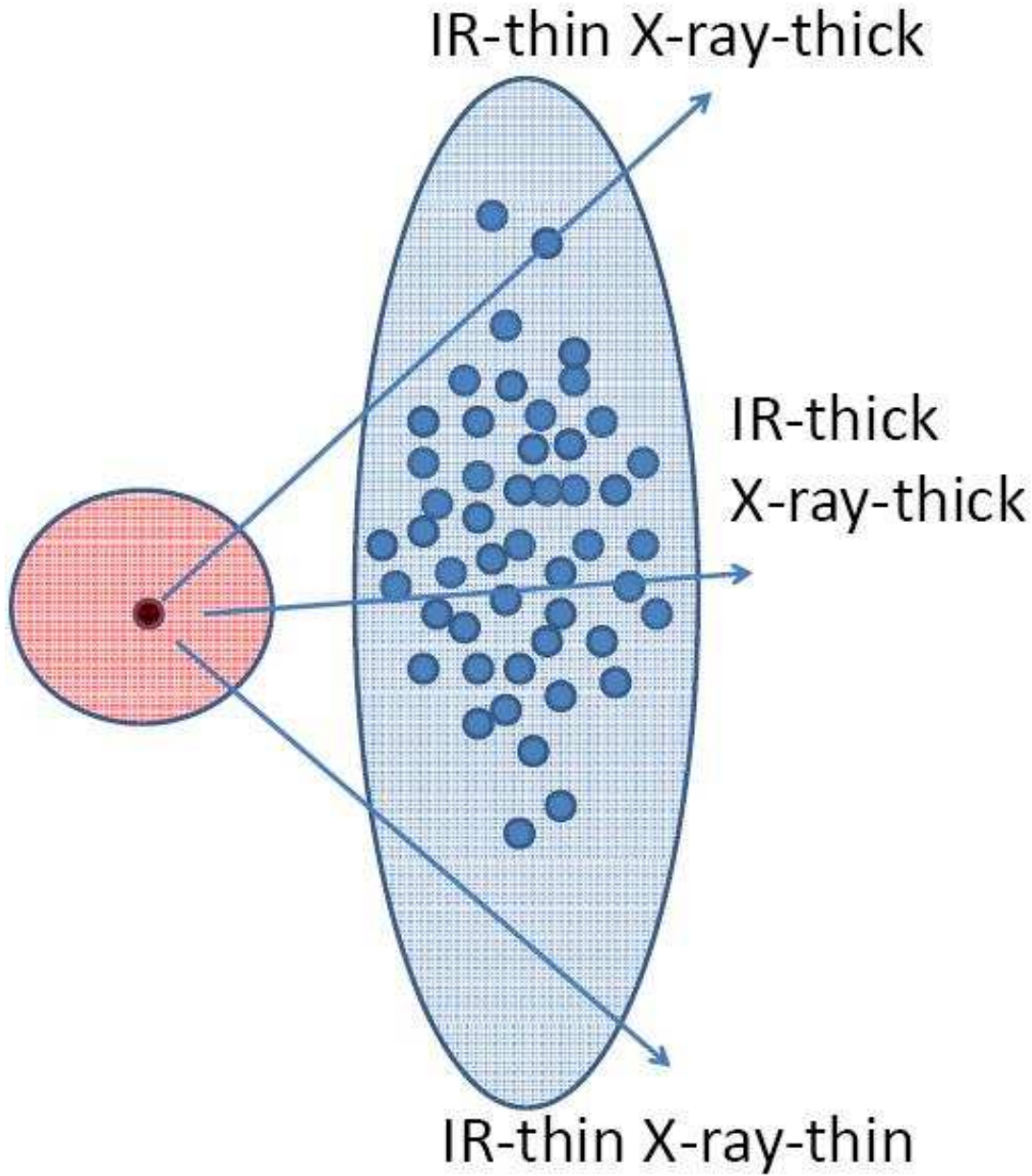}} \caption{A cartoon
diagram demonstrates why we see IR-thin X-ray-thick S2s, and why
they are special.  While we observe the central X-ray source
(plotted as black dot) through the dusty medium (large blue region)
without intervening these Compton-thick clouds (small blue dots), we
see them as IR-thin/X-ray-thin, otherwise as IR-thin/X-ray-thick. In
the situation that the density (or sky coverage) of the
Compton-thick clouds is too high that most of the BLR (red circle)
could be blocked by the Compton-thick clouds, we see them
IR-thick/X-ray-thick. Besides, such three situations might be an
evolving effect, instead of an inclination effect as shown in this
cartoon, i.e., the torus evolves with time, and the appearance of
S2s depends on the evolving status of the torus, instead of
inclination angle. }
\end{center}
\end{figure}

Alternatively, the discrepancy could be to due to the facts that IR and X-ray observations were not obtained simultaneously. For instance, sources with central engines fading off might be mis-classified as Compton-thick against their Compton-thin nature, because the observed X-ray spectra could be dominated by reflection component from a much larger scale due to the fading of the direct emission (e.g. Matt, Guainazzi \& Maiolino 2003).

Both the later two models we presented above predict or require spectral transitions (i.e. from Compton-thin to Compton-thick or vice versa) for our IR-thin X-ray-thick S2s. By searching previous X-ray observations, we found X-ray spectral transitions in two out of six IR-thin X-ray-thick S2s (NGC 1386 and NGC 7674).

Risaliti et al. (2002) have studied NGC 1386 in X-ray and found it
changed from Compton-thin (N$\rm_H$ = 2.8 $\times$ 10$^{23}$
cm$^{-2}$) on Jan 25, 1995 to Compton-thick on Dec 10, 1996, which
means that the timescale of variation is less than 2 years. This
source likely remains Compton-thick till XMM observation in Dec.
2002 (Guainazzi et al. 2005).

Bianchi et al. (2005) reported the hard X-ray flux density of
NGC~7674 decreased from 24 $\times$ 10$^{-12}$ erg cm$^{-2}$
s$^{-1}$ in 1977 (HEAO A-1) to 8 $\times$ 10$^{-12}$ erg cm$^{-2}$
s$^{-1}$ in 1989 (GINGA). BeppoSAX in 1996 and XMM-Newton in 2004
also observed this source, with much smaller hard X-ray flux density
of 0.75 and 0.70 $\times$ 10$^{-12}$ erg cm$^{-2}$ s$^{-1}$,
respectively. Bianchi et al. also presented a spectral transition
between GINGA observations (1989) and BeppoSAX/XMM-Newton
observations, from transmission-dominated to reflection-dominated.
They argued it was likely due to the ``switching-offÓ of the central
engine in NGC~7674, and the reflection-dominated spectra observed by
BeppoSAX and XMM are the delayed emission (with 8$\sim$15 years
delay) from a reflector at larger distance (such as the inner
surface of a torus) although they can not rule out the variation of
absorption column density. Here we argue that the transition is more
likely due to the variation in absorption based on two reasons: 1)
the size of the inner torus radius in Seyfert galaxies is believed
to be within several light weeks to several light months (see
Suganuma et al. 2006), much smaller than that derived based on X-ray
transition ( 8$\sim$15 years); this means the delayed reflection
emission should disappear several months after Óswitching-offÓ and
can not last 8$\sim$15 years; 2) the broad emission line, which is
produced at even smaller distance than the torus and should
disappear even faster after Óswitching-offÓ, was detected in near-IR
6 years after the so-called  ``switching-off'' (Riffel et al. 2006).



No X-ray transitions were reported in the rest four IR-thin X-ray-thick S2s. Note Turner et al. (\cite{T97b}) and Awaki et al. (2000) both reported the variability of hard X-ray flux in Mrk 3 within one year.  More interestingly, recent Suzaku observation (Ikeda, Awaki \& Terashima 2009) suggested Mrk 3 to be viewed at a particular inclination (along the edge  of the torus).
For NGC~6240, no X-ray transition is observed though there was a flux variability of $\sim$
1.7 factor from 1994 to 1998 (Ptak et al. \cite{P03}).   Note NGC 6240 is well known because it contains a binary active galactic nucleus (Komossa et al. 2003).
For Mrk~477 and IRAS 11058-1131, X-ray observations are rather limited. The
available X-ray spectral parameters are from ASCA and BeppoSAX about
20 years ago for Mrk~477 and IRAS 11058-1131, respectively (Bassani et al. 1999; Risaliti et al. 2000).

We finally propose based on our toy model and above observational evidences that IR-thin X-ray-thick S2s are special type of X-ray-thick S2s. They could be viewed at an intermediate angle (i.e., along the edge of Compton-thick torus) comparing with IR-thick X-ray-thick S2s,  and might have higher rate to show spectral transitions than other X-ray thick S2s. We note Guainazzi et al. (2005)  detected only one transition out of 11 optically selected Compton-thick AGNs.
Following up or monitor observations of this particular class of sources could bring valuable informations on the obscuration in AGNs.

\section*{Acknowledgments}
H.C. gratefully acknowledges the support of K.C.Wong Education
Foundation, Hong Kong, and China Postdoctoral
Science Foundation NO.20080430769. The work was supported by Chinese NSF through NSFC10773010/10825312, and the Knowledge Innovation Program of CAS (Grant No. KJCX2-YW-T05).


\begin{thebibliography}{}
\bibitem[Acker et al. (1991)]{A91}Acker, A., Stenholm, B., \& Veron, P. 1991, A\&AS, 87, 499
\bibitem[Alexander et al. (1999)]{A99}Alexander, D. M., Hough, J. H., Young, S., Bailey, J. A., Heisler, C. A., Lumsden, S. L., Robinson, A. 1999, MNRAS,303,L17
\bibitem[ 1985]{A85}Antonucci, R., \& Miller, J. S. 1985, ApJ, 297, 621

\bibitem[ 1993]{A93} Antonucci,
R.R.J. 1993, ARA\&A, 31, 473

\bibitem[ Awaki et al. 2006]{A06}Awaki, H., Murakami, H., Ogawa, Y., Leighly, K. M. 2006, ApJ, 645, 928
\bibitem[Axon et al. (2000)]{Ax00}Axon, D. J., Capetti, A., Fanti, R., Morganti, R., Robinson,
A., Spencer, R. 2000, ApJ, 120, 2284
\bibitem[1999]{B99}Bassani, L., Dadina, M., Maiolino, R., Salvati, M., Risaliti, G., della Ceca, R., Matt, G., Zamorani, G. 1999, ApJS, 121, 473
\bibitem[Bellamy et al. 2003]{B03}Bellamy, M. J., Tadhunter, C. N., Morganti, R., Wills, K. A., Holt, J., Taylor, M. D., Watson, C. A. 2003, MNRAS, 344, L80
\bibitem[Bian 2007]{B07}Bian, W. \& Gu, Q.
2007, ApJ, 657, 159
\bibitem[ 2005]{B05a}Bianchi, S. et al. 2005a, A\&A, 422, 185
\bibitem[ 2005]{B05b}Bianchi, S., Miniutti, G., Fabian, A., Iwasawa, K. 2005b, MNRAS,
360, 380
\bibitem[Bianchi et al. 2008]{B08}Bianchi, S., Corral, A., Panessa, F, Barcons, X, Matt, G. 2008, MNRAS,
385, 195
\bibitem[Bonatto \& Pastoriza 1997]{B97}Bonatto, C. J. \& Pastoriza, M. G.
1997, ApJ, 486, 132

\bibitem[Churazov et al. 2003]{C03} Churazov, E., Forman, W., Jones, C., Bohringer, H. 2003,
ApJ, 590, 225
\bibitem[Clements et al. 2002]{C02}Clements, D. L. et al. 2002, ApJ, 581, 974
\bibitem[Croston et al. 2004]{C04} Croston, J. H. et al. 2004, MNRAS, 353, 879
\bibitem[Dahari \& De Robertis 1988]{D88} Dahari, O., \& De Robertis, M. M. 1988, ApJS, 67, 249
\bibitem[Deluit 2004]{D04}Deluit, S. J. 2004, A\&A, 415, 39
\bibitem[Dennefeld et al. 2003]{D03}Dennefeld, M., Boller, T., Rigopoulou, D., Spoon, H. W. W.
2003, A\&A, 406, 527
\bibitem[DePoy et al. 1987]{D87}Depoy, D. L., Becklin, E. E., Geballe, T.
R. 1987, ApJ, 316, L63
\bibitem[Elitzur \& Ho 2009]{}Elitzur, M. \& Ho, L. C. 2009, ApJ, 701, 91
\bibitem[Evans et al. 2004]{E04}Evans, D. A. et al. 2004, ApJ, 612, 786
\bibitem[Evans et al. 2007]{E07}Evans, D. A. et al. 2007, ApJ, 671, 1345

\bibitem[Franceschini et al. 2003]{F03}Franceschini, A., et al. 2003, MNRAS, 343, 1181
\bibitem[Gonzalez et al. 2001]{G01}Gonzalez Delgado R.M., Heckman T., \& Leitherer C., 2001, ApJ, 546,
845
\bibitem[ 1994]{G94} Goodrich, R. W., Veilleux, S., Hill, G. J. 1994, ApJ, 422, 521
\bibitem[2003]{G03}Gordon, K. D. et al. 2003, ApJ, 594, 279
\bibitem[Gu 2001]{Gu01}Gu, Q., Dultzin-Hacyan, D., de Diego, J.
A. 2001, RMxAA, 37, 3
\bibitem[Guainazzi et al. 2005]{G05}Guainazzi, M., Matt, G., Perola, G.
C. 2005, A\&A, 444, 119

\bibitem[Hardcastle et al. 2006]{H06}Hardcastle, M. J., Evans, D. A., Croston, J.
H. 2006, MNRAS, 370, 1893
\bibitem[ 1997]{H97}Heisler, C. A., Lumsden S. L., \& Bailey J. A., 1997, Nature, 385,
700
\bibitem[ 1999]{H99}Heisler, C. A., De Robertis, M. M. 1999, AJ, 118, 2038

\bibitem[Hill et al. 1996]{H96}Hill, G. J., Goodrich, R. W., Depoy, D.
L. 1996, ApJ, 462, 163
\bibitem[Holt et al. 2006]{H06}Holt, J., Tadhunter, C., Morganti, R., Bellamy, M., Gonz¨¢lez Delgado, R. M., Tzioumis, A., Inskip, K. J. 2006, MNRAS, 370, 1633
\bibitem[]{}Ikeda, S., Awaki, H., \& Terashima, Y. 2009, ApJ, 692, 608
\bibitem[Imanishi \& Terashima 2004]{T04}Imanishi, M., \& Terashima,
Y. 2004, AJ, 127, 758
\bibitem[Itoh et al. 2008]{I08}Itoh, T. et al. 2008, 60, 251, PASJ
\bibitem[]{}Komossa, S. 1999, in ASCA-ROSAT Workshop on AGN, ed. T. Takahashi \&
H. Inoue (ISAS Report 149; Tokyo: ISAS), 149
\bibitem[]{}Komossa, S. et al. 2003, ApJ, 582, L15
\bibitem[Koski 1978]{K78}Koski, A. T. 1978, ApJ, 223, 56
\bibitem[Levenson et al. 2005]{L05}Levenson, N. A., Weaver, K. A., Heckman, T. M., Awaki, H., Terashima, Y.
2005, ApJ, 618, 167
\bibitem[Levenson et al. 2006]{L06}Levenson, N. A., Heckman, T. M., Krolik, J. H., Weaver, K. A., \.{Z}ycki, P. T.
2006, 648, 111
\bibitem[ 2001]{L01}Lumsden, S. L., Heisler, C. A., Bailey, J. A., Hough, J. H., Young, S.
2001, MNRAS, 327, 459
\bibitem[]{}Lutz, D. 1999, in The Universe as seen by ISO, ed. P. Cox, \& M. F. Kessler, ESA-SP427, 623
\bibitem[ 2002]{L02}Lutz, D., Maiolino, R., Moorwood, A. F. M., Netzer, H., Wagner, S. J.
2002, A\&A, 396, 439
\bibitem[Madejskiet et al. 2006]{M06}Madejski, G., Done, C., \.{Z}ycki, P. T., Greenhill, L.
2006, ApJ, 636, 75
\bibitem[]{}Maiolino, R. et al. 2001a, A\&A, 365, 28
\bibitem[]{}Maiolino, R., Marconi, A. \& Oliva, E. 2001b, A\&A, 365, 37
\bibitem[Martin et al. 1983]{M83}Martin, P. G., Thompson, I. B., Maza, J., Angel, J. R.
P. 1983, ApJ, 266, 470
\bibitem[Matt et al. 2003]{}Matt, G., Guainazzi, M., Maiolino, R. 2003, MNRAS, 342, 422
\bibitem[Miller \& Goodrich 1990]{M90}Miller, J. S. \& Goodrich, R. W.
1990, ApJ, 355, 456
\bibitem[Moran et al. 2000]{M00} Moran, E. C., Barth, A. J., Kay, L. E., Filippenko,
A. V. 2000, ApJ, 540, L73
\bibitem[Moran et al. 2001]{M01}Moran, E. C., Kay, L. E., Davis, M., Filippenko, A. V., Barth, A. J.
2001, ApJ, 556, L75
\bibitem[Moran et al. 2007]{M07}Moran, E. C., Barth, A. J., Eracleous, M., Kay, L. E.
2007, ApJ, 668, 31
\bibitem[Mulchaey 1994]{M94} Mulchaey, J. S., Koratkar, A., Ward, M. J., Wilson, A. S., Whittle, M., Antonucci, R. R. J., Kinney, A. L., Hurt, T.
1994, ApJ, 436, 586
\bibitem[Murphy et al. 1999]{M99}Murphy, T. W. Jr., Soifer, B. T., Matthews, K., Kiger, J. R., Armus, L. 1999, ApJ, 525, L85
\bibitem[Murphy et al. 2000]{M00}Murphy, T. W. Jr., Soifer, B. T., Matthews, K., Armus,
L. 2000, AJ, 120, 1675
\bibitem[Nagao et al. 2000]{N00}Nagao, T., Taniguchi, Y., \& Murayama, T.
2000, ApJ, 119, 2605
\bibitem[Nagao et al. 2004]{N04}Nagao, T., Kawabata, K. S., Murayama, T., Ohyama, Y., Taniguchi, Y., Shioya, Y., Sumiya, R., Sasaki, S. S. 2004, AJ,
128, 2066
\bibitem[Nakajima et al. 1991]{N91}Nakajima, T., Kawara, K., Nishida, M., Gregory,
B. 1991, ApJ, 373, 452
\bibitem[Netzer et al. 2005]{N05}Netzer, H., Lemze, D., Kaspi, S., George, I. M., Turner, T. J., Lutz, D., Boller, T., Chelouche,
D. 2005, ApJ, 629, 739
\bibitem[ 2003]{N03}Nicastro, F., Martocchia, A., Matt,
G. 2003, ApJ, 589, L13

\bibitem[Ohno et al. 2004]{O04}Ohno, M., Fukazawa, Y., Iyomoto,
N. 2004, PASJ, 56, 425
\bibitem[Oliva et al. 1998]{O98}Oliva, E., Marconi, A., Cimatti, A., Alighieri, S. D.
1998, A\&A, 329, L21
\bibitem[Piconcelli et al. 2007]{Pi07}Piconcelli, E. et al. 2007, A\&A, 466, 855
\bibitem[Piconcelli et al. 2008]{Pi08}Piconcelli, E., Bianchi, S., Miniutti, G., Fiore, F., Guainazzi, M., Jimenez-Bailon, E., Matt,
G. 2008, A\&A, 480, 671
\bibitem[Pounds \& Vaughan 2006]{P06}Pounds, K., \& Vaughan, S. 2006, MNRAS,
368, 707
\bibitem[Privon et al. 2008]{P08}Privon, G. C., O'Dea, C. P., Baum, S. A., Axon, D. J., Kharb, P., Buchanan, C. L., Sparks, W., Chiaberge,
M. 2008, ApJS, 175, 423
\bibitem[ 2003]{P03}Ptak, A., Heckman, T., Levenson, N. A., Weaver, K., Strickland,
D. 2003, ApJ, 592, 782
\bibitem[]{}Ramos Almeida, C. et al. 2009, ApJ, 702, 1127
\bibitem[Reunanen et al. 2002]{R02}Reunanen J., Kotilainen J. K., \& Prieto, M.
A. 2002, MNRAS, 331, 154
\bibitem[ 2003 ]{R03}Reunanen J., Kotilainen J. K., \& Prieto, M.
A. 2003, MNRAS, 343, 192
\bibitem[ 2006]{R06}Riffel, R., Rodr¨ªguez-Ardila, A., Pastoriza, M.
G. 2006, A\&A, 457, 61
\bibitem[1999]{R99}Risaliti, G., Maiolino, R., Salvati, M.
1999, 522, 157
\bibitem[Risaliti et al. 2000]{R00}Risaliti, G., Gilli, R., Maiolino, R., Salvati, M. 2000, AA,
357, 13
\bibitem[2002]{Ri02}Risaliti, G., Elvis, M., Nicastro,
F. 2002, ApJ, 571, 234
\bibitem[ 2007]{R07}Risaliti G., Elvis M., Fabbiano G., Baldi A., Zezas A. et
    al. 2007, ApJ, 659, L111
\bibitem[Ruiz et al. 1994]{R94}Ruiz, M.; Rieke, G. H.; Schmidt, G.
D. 1994, ApJ, 423, 608
\bibitem[]{}Suganuma, M. et al. 2006, ApJ, 639, 46
\bibitem[Sazonov et al. 2007]{S07}Sazonov, S., Revnivtsev, M., Krivonos, R., Churazov, E., Sunyaev,
    R., 2007, A\&A, 462, 57
\bibitem[ 2007]{Sh07} Shu, X. W., Wang, J. X., Jiang, P., Fan, L. L., Wang, T. G.
2007, ApJ, 657, 167
\bibitem[Storchi-Bergmann \& Pastoriza 1989]{S89}Storchi-Bergmann, T., \& Pastoriza, M. G.
1989, ApJ, 347, 195
\bibitem[Storchi-Bergmann et al. 1995]{S95}Storchi-Bergmann, T., Kinney, A. L., Challis, P.
1995, ApJS, 98, 103
\bibitem[Teng et al. 2005]{T05}Teng, S. H., Wilson, A. S., Veilleux, S., Young, A. J., Sanders, D. B., Nagar, N. M.
2005, ApJ, 633, 664
\bibitem[Teng et al. 2008]{T08}Teng, S. H. et al. 2008, ApJ, 674, 133
\bibitem[Teng et al. 2009]{T09}Teng, S. H. et al. 2009, ApJ, 691, 261
\bibitem[Terlevich et al. 1991]{T91}Terlevich, R., Melnick, J., Masegosa, J., Moles, M., Copetti, M. V. F.
1991, A\&AS, 91, 285
\bibitem[Tran et al. 1992]{T92}Tran, H. D., Miller, J. S., Kay, L. E.
1992, ApJ, 397, 452
\bibitem[Tran 2001]{T01}Tran, H. D. 2001, ApJ, 554, L19;
\bibitem[Turner et al. 1997]{T97a}Turner, T. J., George, I. M., Nandra, K., Mushotzky, R.
F., 1997, ApJ, 113, 23 (1997a)
\bibitem[ 1997]{T97b}Turner, T. J., George, I. M., Nandra, K., Mushotzky, R.
F. 1997, ApJ, 488, 164 (1997b)


\bibitem[Ueno et al. 2000]{U00}Ueno, S.; Ward, M. J.; O'Brien, P. T.; Stirpe, G. M.; Matt,
G. 2005, AdSpR, 25, 823
\bibitem[Vaceli et al. 1997]{Va97}Vaceli, M. S., Viegas, S. M., Gruenwald, R., de Souza, R. E.
1997, AJ, 114, 1345

\bibitem[Veilleux et al. 1995]{V95}Veilleux, S., Kim, D.-C., Sanders, D. B., Mazzarella, J. M., Soifer, B. T.
1995, ApJS, 98, 171
\bibitem[ 1997a]{V97a}Veilleux, S., Goodrich, R. W., \& Hill, G.J. 1997, ApJ,
477,631 (1997a)
\bibitem[Veilleux et al. 1997]{V97b}Veilleux, S., Sanders, D. B., \&
Kim, D. -C. 1997, ApJ, 484, 92 (1997b)

\bibitem[Veilleux et al. 1999]{V99}Veilleux Sylvain, Sanders D. B. \& Kim, D.-C.
1999, ApJ, 522, 139
\bibitem[]{}Wang, T. G. et al. 2009, AJ, 137, 4002
\bibitem[]{}Wang, J. M. \& Zhang, E. P. 2007, ApJ, 660, 1072
\bibitem[Warwick et al. 1993]{W93}Warwick, R. S., Sembay, S., Yaqoob, T., Makishima, K., Ohashi, T., Tashiro, M., Kohmura,
Y. 1993, MNRAS, 265, 412
\bibitem[Xia et al. 2002]{X02}Xia, X. Y., Xue, S. J., Mao, S., Boller, Th., Deng, Z. G., Wu, H. 2002,
ApJ, 564, 196
\bibitem[Young et al. 1996]{Y96}Young, S., Hough, J. H., Efstathiou, A., Wills, B. J., Bailey, J. A., Ward, M. J., Axon, D. J.
1996, MNRAS, 281, 1206
\bibitem[Young et al. 1998]{Y98}Young, S., Hough, J. H., Axon, D. J., Fabian, A. C., Ward, M.
J. 1998, MNRAS, 294, 478
\bibitem[Zhang 2006]{Z06}Zhang, J. S., Henkel, C., Kadler, M., Greenhill, L. J., Nagar, N., Wilson, A. S., Braatz, J. A. 2006,
A\&A, 450, 933


\end{thebibliography}
\end{document}